\documentclass[conference]{IEEEtran}
\IEEEoverridecommandlockouts
\usepackage{cite}
\usepackage{amsmath,amssymb,amsfonts}
\usepackage{algorithmic}
\usepackage{graphicx}
\usepackage{textcomp}
\usepackage{xcolor}
\usepackage{units}
\usepackage{multirow}
\usepackage{colortbl}
\usepackage{soul}
\usepackage{ragged2e}
\usepackage[left=0.625in, right=0.625in,top=0.75in,bottom=1in]{geometry}

\usepackage{amsthm}
\usepackage{thmtools, thm-restate}
\usepackage{algorithm}

\usepackage{hyperref}
\def\BibTeX{{\rm B\kern-.05em{\sc i\kern-.025em b}\kern-.08em
    T\kern-.1667em\lower.7ex\hbox{E}\kern-.125emX}}
\begin{document}

\title{Semantic Text Compression for Classification
%\thanks{This work is supported in part by NSF CNS-2112471.}
}

\author{Emrecan Kutay and Aylin Yener
\\ INSPIRE@OhioState Research Center 
\\Dept. of Electrical and Computer Engineering
\\ The Ohio State University
\\ kutay.5@osu.edu, yener@ece.osu.edu 
}

\newgeometry{left=0.625in, right=0.625in,top=0.75in,bottom=1in}

\maketitle

\begin{abstract}
We study semantic compression for text where meanings contained in the text are conveyed to a source decoder, e.g., for classification. The main motivator to move to such an approach of recovering the meaning without requiring exact reconstruction is the potential resource savings, both in storage and in conveying the information to another node. Towards this end, we propose semantic quantization and compression approaches for text where we utilize sentence embeddings and the semantic distortion metric to preserve the meaning. Our results demonstrate that the proposed semantic approaches result in substantial (orders of magnitude) savings in the required number of bits for message representation at the expense of very modest accuracy loss compared to the semantic agnostic baseline. We compare the results of proposed approaches and observe that resource savings enabled by semantic quantization can be further amplified by semantic clustering. Importantly, we observe the generalizability of the proposed methodology which produces excellent results on many benchmark text classification datasets with a diverse array of contexts.

\end{abstract}

\begin{IEEEkeywords}
Semantic communications, sentence embedding, semantic text compression. 
\end{IEEEkeywords}

\section{Introduction}
\label{sec:intro}
6G is expected to bring human and machine communication closer together, and more and more over wireless resource constrained links \cite{6gsurvey}. To this end, efficient representation and communication of human-like information exchange between network nodes is a key pillar of the 6G vision \cite{6gandbeyond}.

Conventional compression and communication designs abstract the process of information exchange as conveying digital sequences to their end points reliably and efficiently, in a content-agnostic fashion \cite{Shannon}. This is a key-assumption in the landmark framework by Shannon, and is a powerful abstraction that arguably single-handedly ushered in the digital era that we enjoy today. Whereas Shannon's framework identified fundamental limits of digital communications, which have been crucial in building communication networks, next generation systems and future use cases are becoming increasingly complex, often beyond just needing communications, but {\it information that has a purpose}. Many of these require stringent latency and reliability requirements making them resource intensive, e.g., smart cities, and fully autonomous vehicles, and are limited by the capacity of current models. Furthermore, human-driven networks and human-machine interaction are becoming more pervasive driving all future applications. These point out to a need toward considering content and semantic (meaning)- aware models of next generation systems, making semantic communications a meaningful direction forward \cite{beyondtransmittingbits}.
 
Reference \cite{guler2014semantic} is the first study that introduces semantic communications, i.e., communicating with the goal of conveying the meaning. Specifically, the reference introduces the concept of “semantic distortion” as a metric instead of conventional syntactic metrics like symbol error rate (SER), and demonstrates the potential of resource savings when only meanings are meant to be conveyed as opposed to the exact reconstruction of messages. Building on this novel metric, reference \cite{semcom_game} studies a game theoretic approach towards semantic communication with an external agent of probabilistic nature that can provide helpful or adversarial context to the receiver. Optimal transmission policy and nash equilibria are investigated for the agent and player whose aim is to optimize the encoder-decoder structure. Utilizing knowledge graphs (KG), the concept of semantic compression is studied in \cite{guler2013_semanticcompression} by investigating upper and lower bounds on codeword length. Reference \cite{guler2017_entropy} further builds upon this direction for interactive fact building (function computation) between two nodes with potentially differing knowledge bases. 

 Recent efforts utilize neural networks (NN) for joint source channel coding, and provide a pathway toward moving away from the separation of source and channel coding which is practical in model based designs, but is not necessarily optimal (e.g. in finite blocklengths). Reference \cite{toshea_dl} shows that replacing the conventional communication blocks with NN and performing end-to-end optimization with an autoencoder structure results in a better bit error rate (BER) compared to traditional baselines like uncoded binary phase shift keying (BPSK) and hamming code. Using a similar framework, \cite{zhijin_deepsc} has proposed a model for text communication conveying the semantic meaning, focusing on developing specialized models for given background knowledge. Further studies are conducted for different modalities like image, speech, and video transmission where attention-based models like squeeze and excitation (SE), and convolutional neural networks (CNN), are used to extract semantic information in parallel to the communication task \cite{beyondtransmittingbits}. All these models require a re-design and re-training for different background knowledge and thus are specialized for each application/data set. Building on these previous efforts, but aiming for generalizability, in this paper we investigate how to communicate more efficiently by conveying the meaning of a message, for a broad range of textual contexts.

We define a distortion metric for text to quantify the amount of semantic error and propose generalizable semantic quantization and compression approaches to minimize this distortion. In doing so, we aim to significantly reduce the number of bits required in message representation. Proposed approaches are able to extract semantics by sentence embeddings where a corresponding vector in \begin{math} R^p \end{math} is obtained for each message. Additionally, we argue that  compression in semantic quantization can be improved by utilization of clustering in the semantic space, i.e., clustering of text embeddings. Results on benchmark text classification datasets show that there is a significant improvement in efficiency with only a modest decrease in classification accuracy. Our approaches do not rely on specifics of the topic (background), and re-training is not required for different contexts. We obtain results using the same parameters and design for all datasets unless specifically stated. The main contributions of this paper can be stated as follows:

\begin{itemize}
    \item We propose DL-based generalizable semantic text quantization and compression approaches for text classification, where classification is performed via proposed models utilizing sentence embeddings and clustering in the semantic space. 
    \item Simulation results demonstrate that proposed approaches reduce the number of bits in message representation by orders of magnitude compared to the semantic agnostic conventional approach at the expense of a modest decrease in accuracy. Comparing the proposed approaches, semantic compression provides further savings with approximately the same error rate.
    \item We demonstrate the knowledge-independent structure of the proposed methodology through multiple datasets with different contexts. Results show that approaches are able to extract semantics and provide efficiency without any re-training or re-design, validating generalizability to any background knowledge.
\end{itemize}

\section{Background}
\label{sec:background}
In this section, we provide a brief background for sentence embeddings \cite{sentence-bert} and affinity propagation (AP) \cite{affinity_prop} relevant to our proposed approaches.
\subsection{Sentence Embeddings}
Sentence - bidirectional encoder representations from transformers (SBERT) is an extension of the state-of-the-art BERT model providing a fixed-length embedding vector in \begin{math} R^p\end{math} for a sentence or short paragraph. The purpose is to map semantically similar sentences to closer and dissimilar ones to distant vectors, enabling semantic search and clustering for textual content \cite{sentence-bert}. 

BERT is a transformer-based natural language processing (NLP) model that first initialized the bi-directionality in language models allowing usage of both left and right contexts \cite{bert}. As an example of uni-directionality, the left-to-right architecture allows the current token to only depend on previous ones. However, BERT uses the masked language model objective, i.e., randomly masking some of the tokens and trying to predict them from the contexts on both sides, providing bi-directionality. This structure set a state-of-the-art performance level for NLP tasks, i.e., language understanding, question-answering, and sentence classification. There are studies to obtain fixed-size sentence embeddings utilizing the classic BERT model by averaging the output layer and using the [CLS] token, but none of them are proven to be usable \cite{sentence-bert}.

SBERT includes an additional pooling layer to classical BERT to provide fixed-size embeddings. In addition to classification and regression objectives, the fundamental concept in SBERT is to include the triplet network concept initialized in  \cite{facenet_triplets}, where there is a positive (p) and negative (n) instance for each sample (a). In SBERT, positive and negative instances correspond to semantically similar and different sentences from training data, respectively. Introducing triplets of data, the distance between positive instant and current sample is tried to be minimized while the distance between negative sample is tried to be maximized. In \cite{sentence-bert}, the triplet loss is formulated as follows:

\begin{equation}
    \label{triplet_loss}
    max(\lVert s_a - s_p \rVert - \lVert s_a - s_n \rVert + \epsilon, 0)
\end{equation}
\noindent
where \begin{math} \lVert . \rVert \end{math} is a distance metric, \begin{math} \epsilon \end{math} is the margin, and \begin{math} s_a, s_p, s_n\end{math} are vector embeddings of a, p, n, respectively \cite{sentence-bert}.

\subsection{Affinity Propagation}
Affinity propagation (AP) is a clustering algorithm with centroids selected from actual samples, making them “exemplars”. The main objective is to cluster sample (i) to exemplar (k) while maximizing the given similarity metric. If the goal is to minimize the \begin{math} \mathit L2 \end{math} distance, the similarity \begin{math} s(.,.)\end{math} between points \begin{math} x_i\end{math} and \begin{math} x_j\end{math} is chosen as follows \cite{affinity_prop}:
\begin{equation}
    \label{aff_prop_similarity}
    s(x_i, x_j) = - \rVert x_i - x_j \lVert_2 ^2
\end{equation} 

Algorithm performs clustering through exchanged "responsibility" and "availability" messages between data points. The “responsibility” messages \begin{math} r(i,k)\end{math} are sent from each sample \textit{i} to candidate exemplar \textit{k}, indicating the score of being an actual exemplar for \textit{i}. The “availability” messages \begin{math} a(i,k)\end{math} are sent from each candidate exemplar \textit{k} to sample \textit{i}, indicating how suitable for sample \textit{i} to choose \textit{k} as an exemplar. Equations for these messages are given as follows \cite{affinity_prop}:

\begin{equation}
    \label{aff_prop_responsability}
    r(i,k) \leftarrow s(i,k) - \max_{k' s.t. k'\neq k} \{ a(i,k') + s(i,k') \}
\end{equation}

\begin{equation}
    \label{aff_prop_availability}
    a(i,k) \leftarrow \min \{ 0, r(k,k) +  \sum_{i' s.t. i' \notin \{i, k\}}^{} max\{0,r(i',k) \} \}
\end{equation}

Algorithm initializes zero availabilities, and responsibilities are set to similarities between points \textit{i} and \textit{k} which is taken as input. The superiority of this algorithm comes from the structure that does not need the number of clusters as an input. Instead, that number is determined by the given input similarities and messaging procedure. The number of clusters can be increased by increasing the input similarity and vice-versa, while a moderate number is obtained by selecting input similarity as the median of the sample similarities \cite{affinity_prop}.

\section{System Model}
\label{sec:system_model}
%We consider text compression for classification where the task is performed at the decoder side using the training dataset, available only at the decoder side, with text embeddings \begin{math} Q_{train}\end{math} and their corresponding labels \begin{math} y_{train} \end{math}. We assume both encoder and decoder calculate and store the embeddings of last \textit{N} messages in matrix \begin{math} Q_{pre}^N\end{math} with dimensions \begin{math} \textit{N}  \times  \textit{p} \end{math}, where \textit{p} is the dimension of embeddings calculated by SBERT, and \begin{math}Q_{pre}^N(i)_{0 \le i < N}\end{math} corresponds to the embedding of \begin{math} i^{th}\end{math} encoded-decoded sentence. The input of proposed approaches is a text \begin{math} \mathbf{s} \end{math}, not necessarily a single sentence, and we define the semantic distance between texts \begin{math} \mathbf{s}(k) \end{math} and \begin{math} \mathbf{s}(j) \end{math} as follows.

We consider text compression for classification where the task is performed at the decoder side using the training dataset, available only at the decoder side, with text embeddings \begin{math} Q_{train}\end{math} and their corresponding labels \begin{math} y_{train} \end{math}. We assume both encoder and decoder have a memory \begin{math} Q_{pre}^N\end{math}, where embeddings of \textit{N} previously encoded-decoded messages are stored. The memory, \begin{math} Q_{pre}^N\end{math}, is a matrix with dimensions \begin{math} \textit{N}  \times  \textit{p} \end{math}, where \textit{p} is the dimension of embeddings calculated by SBERT, and \begin{math}Q_{pre}^N(i)_{0 \le i < N}\end{math} is the embedding of \begin{math} i^{th}\end{math} encoded-decoded message. The input of proposed approaches is a text \begin{math} \mathbf{s} \end{math}, not necessarily a single sentence, and we define the semantic distance between texts \begin{math} \mathbf{s}(k) \end{math} and \begin{math} \mathbf{s}(j) \end{math} as follows:

\begin{equation}
    \label{semantic_distance}
    \Delta(\mathbf{s}(k), \mathbf{s}(j)) = \rVert Q(k) - Q(j) \lVert_2 ^2
\end{equation}
\noindent

\noindent
Where \begin{math} Q(k)\end{math} and \begin{math} Q(j)\end{math} are embeddings of \begin{math} \mathbf{s}(k) \end{math} and \begin{math} \mathbf{s}(j) \end{math}.

\subsection{Conventional Approach}
\label{conventional_model_section}
\begin{figure}[H]
\centering
\includegraphics[width=1.95in]{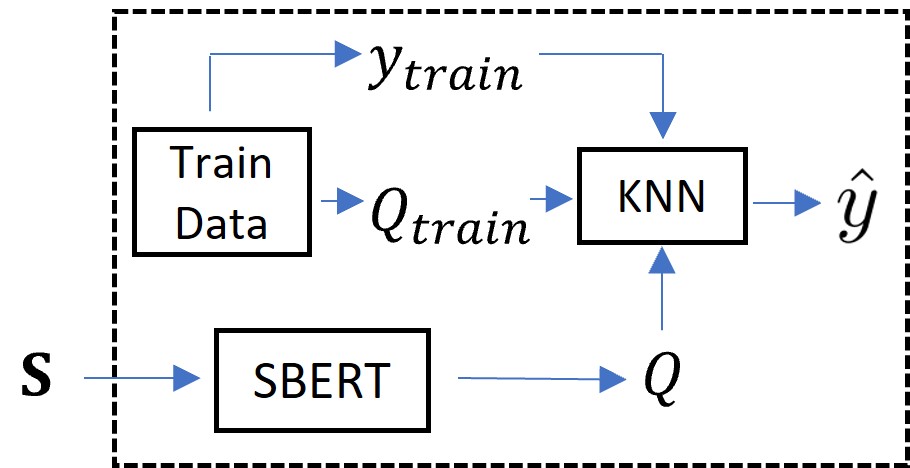}
\caption{The model of conventional approach.}
\label{fig:Conventional_Model} 
\end{figure}

The conventional approach relies on the exact reconstruction of text \begin{math} \mathbf{{s}} \end{math} at the decoder side via Huffman coder and decoder. As shown in fig.\ref{fig:Conventional_Model}, the embedding of \begin{math} \mathbf{{s}} \end{math} is calculated via SBERT, and prediction \begin{math} \hat{y}\end{math} is made by the K-nearest neighbors (KNN) algorithm in the semantic space using \begin{math}Q_{train}\end{math} and \begin{math} y_{train} \end{math}.

We used block coding for a fair comparison. For block size \textit{K}, coding is performed over symbols each containing a sequence of \textit{K} characters.

\subsection{Semantic Quantization Approach} 
\label{semantic_comm_model_section}
\begin{figure}
\centering
\includegraphics[width=2.75in]{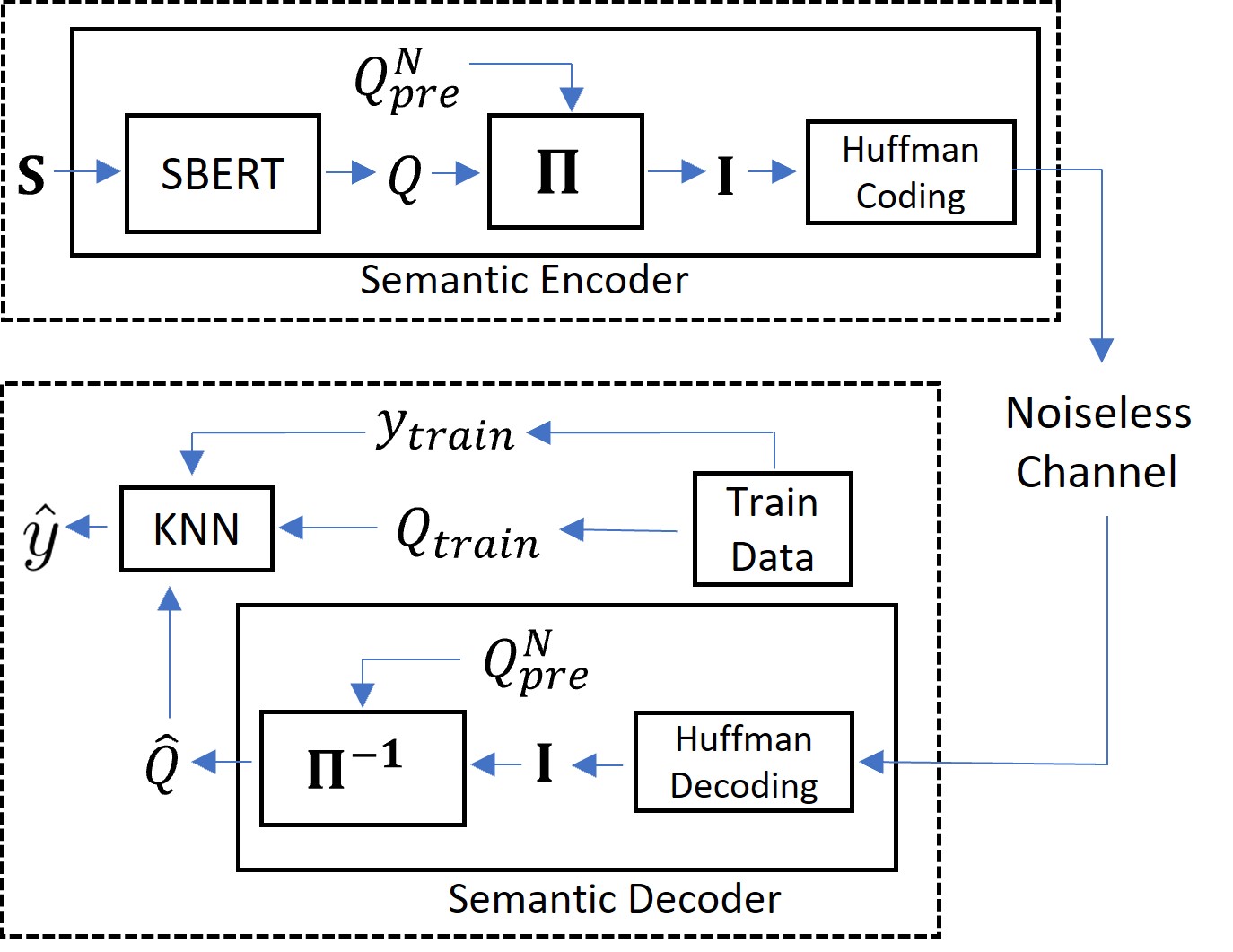}
\caption{The model of proposed semantic quantization approach.}
\label{fig:Baseline_Model} 
\end{figure}

 We propose the semantic quantization approach to eliminate the need for an exact reconstruction of messages. This approach uses \begin{math} Q_{pre}^N\end{math} as the common knowledge between encoder and decoder sides where each embedding in \begin{math} Q_{pre}^N\end{math} is used as a distinct symbol for coding. For this purpose, we define semantic distortion \begin{math} \delta(.,.)\end{math} for text \textbf{s}, when prediction \begin{math} \hat{y}\end{math} is obtained by using \begin{math} \hat{Q}\end{math} in \begin{math} Q_{pre}^N\end{math}, instead of actual embedding \begin{math} Q\end{math} of \textbf{s}. 
\begin{equation}
    \label{semantic_distortion}
    \delta(Q, \hat{Q}) = \rVert Q - \hat{Q} \lVert_2
\end{equation}
\noindent
We define semantic index assignment \begin{math} \mathbf{\Pi}_{Q_{pre}^N}( . )\end{math} for text \begin{math} \mathbf{s}\end{math} with embedding \begin{math} Q\end{math} to minimize eq.\ref{semantic_distortion} as follows:
\begin{equation}
    \label{index_assignment}
    \mathbf{\Pi}_{Q_{pre}^N}( Q \mathbf{)} = \underset{i' s.t. 0\le i' < N}{\mathrm{argmin}} \delta\Bigl(Q, Q_{pre}^N(i')\Bigl)
\end{equation}
As shown in fig.\ref{fig:Baseline_Model}, Huffman coding is performed for index \begin{math} \mathbf{I}(\mathbf{s})\end{math} obtained for text \begin{math} \mathbf{s}\end{math} by  \begin{math}
    \mathbf{I}(\mathbf{s}) = \mathbf{\Pi}_{Q_{pre}^N}( Q \mathbf{)}\end{math}. At the decoder side, decoded index is used to select the corresponding embedding \begin{math} \hat{Q}\end{math} via \begin{math}  \mathbf{\Pi(.)}_{Q_{pre}^N}^{-1}\end{math}. Prediction is then obtained with \begin{math} \hat{Q}\end{math} which is given to KNN along with \begin{math}Q_{train}\end{math} and \begin{math} y_{train} \end{math}.

\subsection{Semantic Compression Approach}
\label{semantic_compression_model_section}

\begin{figure}[H]
\centering
\includegraphics[width=2.75in]{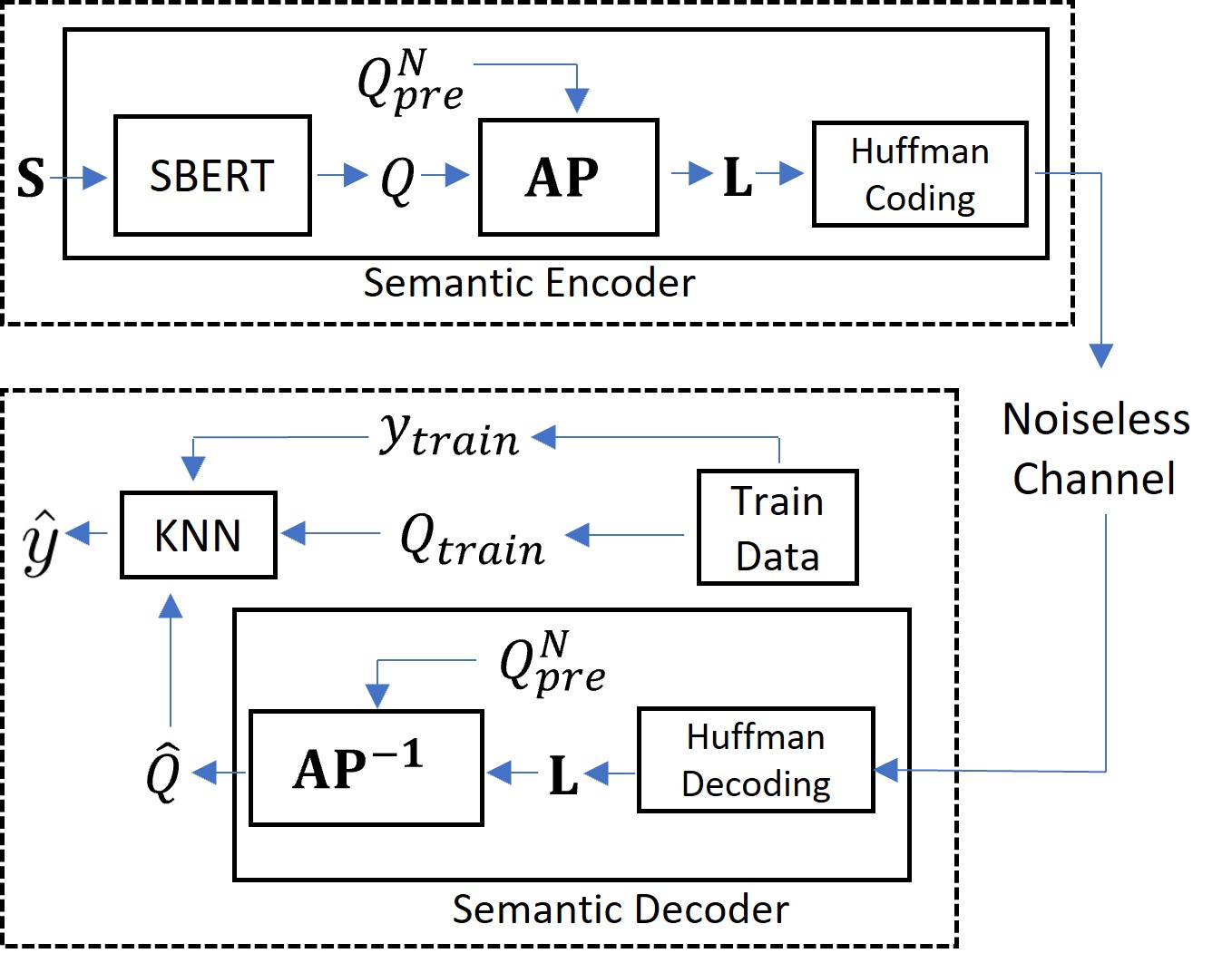}
\caption{The model of proposed semantic compression approach.}
\label{fig:Proposed_Model} 
\end{figure}

We propose the semantic compression approach with the aim of further efficiency compared to the previous section \ref{semantic_comm_model_section} by clustering the stored embeddings in matrix \begin{math} Q_{pre}^N\end{math}. This operation decreases the cardinality of the symbol set in section \ref{semantic_comm_model_section}, which is \textit{N}, by taking cluster labels as symbols instead of taking them as each embedding separately. To this end, we utilize affinity propagation (AP) by converting the proposed semantic distortion eq.\ref{semantic_distortion} to similarity metric in eq.\ref{aff_prop_similarity}. This structure provides generalizability by not requiring any assumption or prior information about the source, i.e., the number of clusters. As presented in fig.\ref{fig:Proposed_Model}, Huffman coding is performed for the cluster label \begin{math} \mathbf{L}(\mathbf{s})\end{math} assigned to text \begin{math} \mathbf{s} \end{math}.

\begin{equation}
    \mathbf{L}(\mathbf{s}) = \mathbf{AP}_{Q_{pre}^N}(Q)
\end{equation}

\noindent
Where \begin{math} Q \end{math} is the embedding of text \begin{math} \mathbf{s} \end{math}, and \begin{math}\mathbf{AP}_{Q_{pre}^N}\end{math} is the AP where clusters are formed over embeddings in \begin{math} Q_{pre}^N \end{math}. Algorithm requires the same random seed in encoder and decoder sides to match the AP structure providing the same number of clusters and exemplars. Corresponding exemplar \begin{math} \hat{Q}\end{math} is selected by the AP at the decoder side, \begin{math} \mathbf{AP}_{Q_{pre}^N} ^{-1}\end{math}, using the decoded cluster label and given to KNN along with \begin{math}Q_{train}\end{math} and \begin{math} y_{train} \end{math}.

\section{Results}
\label{sec:results}
This section provides performance, generalizability analysis, and comparison of approaches proposed in sections \ref{conventional_model_section}, \ref{semantic_comm_model_section}, and \ref{semantic_compression_model_section} on multiple benchmark text classification datasets. Effects of parameters, namely, \textit{N} value and pre-trained neural network in SBERT are observed, and trade-offs are discussed.

\subsection{Datasets and Simulation}
We obtain results through 5 datasets, namely AG's News, DBPedia 14, Humor Detection, IMDB Reviews, and Yelp Polarity \cite{text_classification_dataset_lecun}, \cite{dataset_humor}, \cite{imdb_dataset}. Statistics of the used datasets are given in table \ref{datasets} with the following features: i) Type: topic classification or sentiment analysis; ii) Number of classes; iii) Average length: Number of words per sample (without any preprocessing); iv) Size of the train set. A fair comparison of approaches is made by restricting the size of the test set to 2000, making the remaining variables only the statistics and contexts of datasets. Balanced test sets, in terms of labels, enabled evaluation via accuracy without a need for additional metrics.

\begin{table}[H]
\caption{Dataset Features}
\begin{center}
\begin{tabular}{|c|c|c|c|c|}
\hline
\textbf{Dataset} & \textbf{Type} & \textbf{Classes} & \textbf{Av. Len.} & \textbf{Train Set Sz.} \\
\hline
AG's News & Topic  &  4 & 36 & 50000\\
\hline
DBPedia 14 & Topic  &  14 & 51 & 560000\\
\hline
Humor Detection\textsuperscript{*} & Sentiment  &  2 & 14 & 200000\\
\hline
IMDB Reviews\textsuperscript{**} & Sentiment  &  2 & 265 & 50000\\
\hline
Yelp Polarity & Sentiment & 2 & 156 & 560000\\
\hline
\end{tabular}

\vspace{0.1cm}
 \justify 
 \textsuperscript{*} Test set is created from the training set since a separate test set is not provided; \textsuperscript{**} Due to the small size of this dataset, train and test sets, each having 25000 samples, are processed as a single train set which test samples are then separated from. For the remaining datasets, test samples are chosen from provided test sets.  

\label{datasets}
\end{center}
\end{table}

\noindent 

In this section, results are taken by parameters: \textit{N} is 20000; eq.\ref{aff_prop_similarity} is the similarity for AP; \textit{all-mpnet-base-v2} is SBERT pre-trained neural network; \textit{K} is 15 in KNN. We assumed that a sufficient number of embeddings from each class is stored in \begin{math}Q_{pre}^N\end{math} for semantic index assignment and semantic clustering.

\begin{figure}
\centering
\includegraphics[width=3.4in]{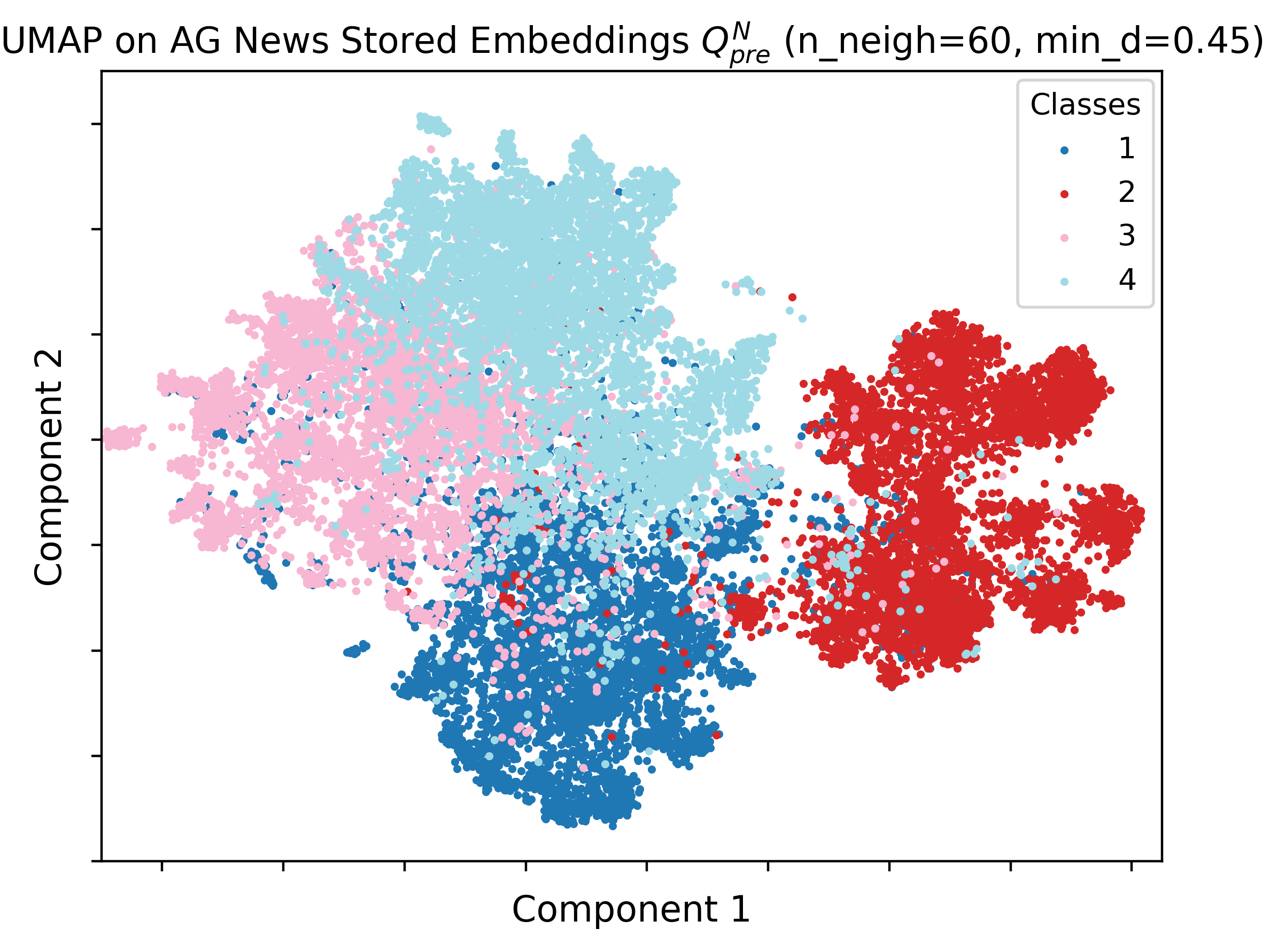}
\caption{2D Visualization of AG News stored embeddings via UMAP.}
\label{fig:AGNews_umap} 
\end{figure}

\begin{figure}[htbp]
\centering
\includegraphics[width=3.4in]{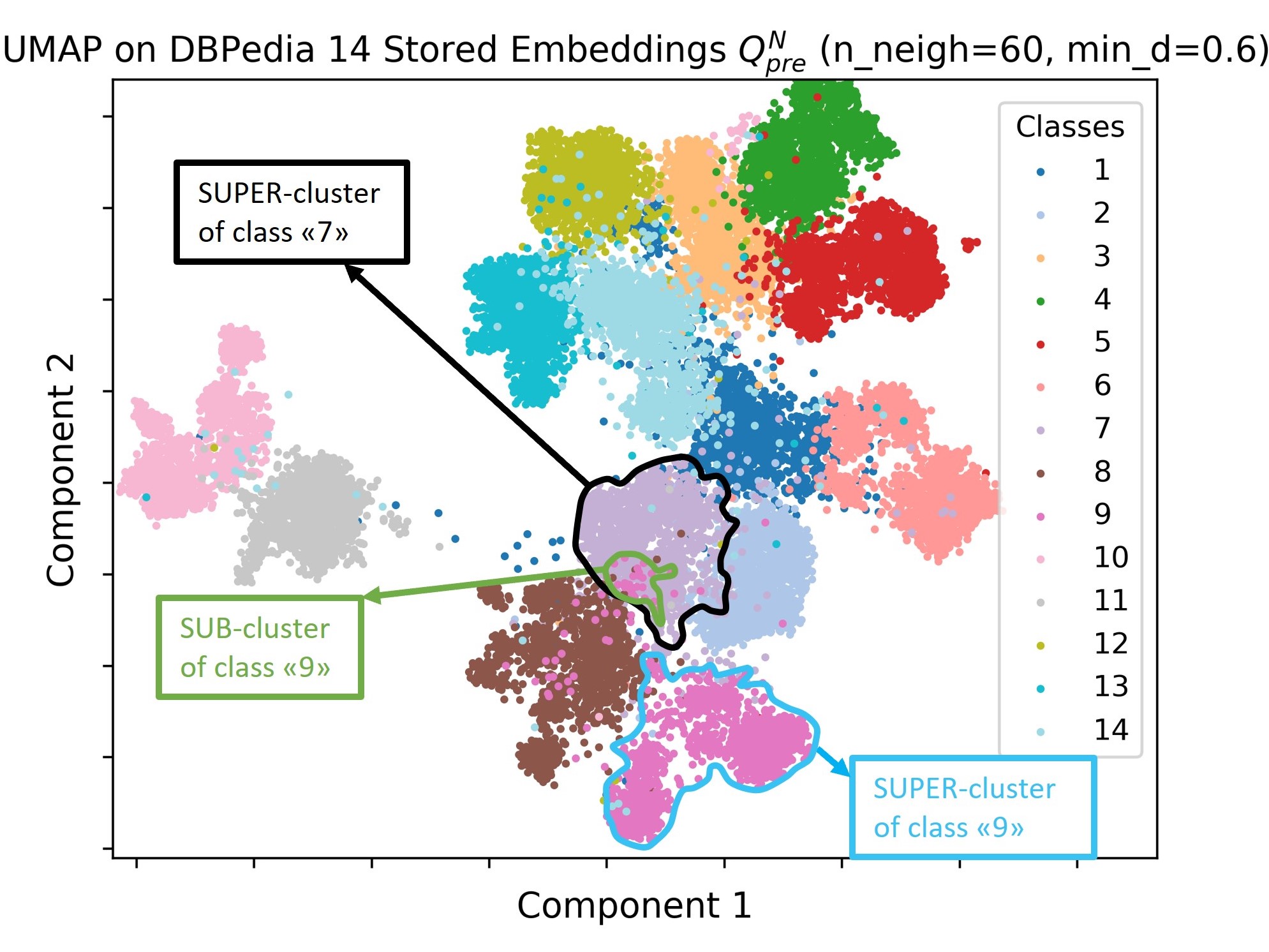}
\caption{2D Visualization of DBPedia 14 stored embeddings via UMAP.}
\label{fig:DBPedia_umap} 
\end{figure}

We use uniform manifold approximation and projection for dimension reduction (UMAP) to project embeddings in \begin{math}Q_{pre}^N\end{math} to \begin{math} R^2 \end{math} \cite{UMAP}. We then observe the distribution of embeddings with respect to their labels for AG's News and DBPedia 14 datasets, which are argued to provide comprehensive results due to their large number of classes compared to other datasets. Distinct clusters for each label in figures \ref{fig:AGNews_umap} and \ref{fig:DBPedia_umap} support the proposed semantic compression since texts with not only similar meanings but also the same labels are projected to nearby vectors. In table \ref{assignedsentence}, we select a random sample from AG's News dataset, stated as \textbf{s}, and obtain messages corresponding to index and cluster label assigned by proposed semantic index assignment \begin{math}\mathbf{I}(\mathbf{s})\end{math} and semantic clustering \begin{math}\mathbf{L}(\mathbf{s})\end{math}, respectively. We compare the assigned messages and conclude that they are consistent with human understanding. Semantics required for the classification task is conveyed while enhancing efficiency, which is the aim of this paper.

\begin{table}[htbp]
\caption{Example of Assigned Messages using AG's News Dataset}
\begin{center}
\begin{tabular}{|c|c|}
\hline
\textbf{Approach} & \textbf{Corresponding Message} \\

\hline 
\begin{math} \mathbf{s} \end{math} & Linksys will provide broadband-to-phone\\
  &   adapters and, eventually, Wi-Fi equipment.\\

\hline 
 & Linksys and Netgear, providers of home network,\\
\begin{math}\mathbf{I}(\mathbf{s})\end{math}  &  are expected to announce they are entering business\\
 &of making gear to place telephone calls over the Internet.\\

\hline 
\begin{math}\mathbf{L}(\mathbf{s})\end{math}& Vonage teams with Wi-Fi equipment makers \\
  &  Linksys and Netgear on voice over Internet Protocol.\\

\hline
\end{tabular}
\label{assignedsentence}
\end{center}
\end{table}

Tables \ref{agnewsresults}, \ref{dbpediaresults}, \ref{humorresults}, \ref{imdbresults}, and \ref{yelpresults} represent the number of bits used in Huffman coding and classification accuracies. Compared to conventional approach, a significant decrease in required bits is observed in proposed approaches without major accuracy degradation. Semantic compression approach achieves approximately the same error rate as semantic quantization with fewer bits which demonstrates that it further compresses the source without losing any meaning required for classification.

The number of clusters extracted by AP, given in the last row of tables, is higher than the number of classes. In addition to high dimensionality and low similarities within the same class, we argue that the main reason is the sub-clusters observed in figures \ref{fig:AGNews_umap}, and \ref{fig:DBPedia_umap} that lay in a different class's super-cluster as represented in fig.\ref{fig:DBPedia_umap}. Although DBPedia 14 dataset has more classes than AG's News dataset, the number of extracted clusters in table \ref{agnewsresults} is much higher than in table \ref{dbpediaresults}. This is consistent with the proposed argument as sub-clusters are more frequent in fig.\ref{fig:AGNews_umap} than fig.\ref{fig:DBPedia_umap} resulting in more clusters.

\begin{table}[H]
\caption{AG News Dataset Results}
\begin{center}
\begin{tabular}{|c|c|c|}
\hline
\textbf{Approach} & \textbf{Number of Bits} & \textbf{Accuracy \%} \\
\hline
Conventional & 1822443  &  89.75\\
\hline
Conventional - Size 2 & 1642754  &  89.75\\
\hline
Conventional - Size 3 & 1482159  &  89.75\\
\hline
Semantic Quantization & 28701  &  88.75\\
\hline
Semantic Compression - 2107 & 21729  &  88.10\\
\hline
\end{tabular}
\label{agnewsresults}
\end{center}
\end{table}

\begin{table}[H]
\caption{DBPedia 14 Dataset Results}
\begin{center}
\begin{tabular}{|c|c|c|}
\hline

\textbf{Approach} & \textbf{Number of Bits} & \textbf{Accuracy \%} \\
\hline
Conventional & 2694268  &  96.60\\
\hline
Conventional - Size 2 & 2394019  &  96.60\\
\hline
Conventional - Size 3 & 2169810  & 96.60\\
\hline
Semantic Quantization & 28721  &  91.10\\
\hline
Semantic Compression - 1380 & 20522  &  90.60\\
\hline
\end{tabular}
\label{dbpediaresults}
\end{center}
\end{table}

\begin{table}[H]
\caption{Humor Detection Dataset Results}
\begin{center}
\begin{tabular}{|c|c|c|}
\hline

\textbf{Approach} & \textbf{Number of Bits} & \textbf{Accuracy \%} \\
\hline
Conventional & 607220  &  93.15\\
\hline
Conventional - Size 2 & 557564  &  93.15\\
\hline
Conventional - Size 3 & 510452  & 93.15\\
\hline
Semantic Quantization & 28725  &  88.75\\
\hline
Semantic Compression - 1784 & 21342  &  85.55\\
\hline
\end{tabular}
\label{humorresults}
\end{center}
\end{table}

\begin{table}[H]
\caption{IMDB Reviews Dataset Results}
\begin{center}
\begin{tabular}{|c|c|c|}
\hline

\textbf{Approach} & \textbf{Number of Bits} & \textbf{Accuracy \%} \\
\hline
Conventional & 11342093  &  85.35\\
\hline
Conventional - Size 2 & 10182680  &  85.35\\
\hline
Conventional - Size 3 & 9224617  & 85.35\\
\hline
Semantic Quantization & 28719  &  80.75\\
\hline
Semantic Compression - 1562  & 20098  &  78.30\\
\hline
\end{tabular}
\label{imdbresults}
\end{center}
\end{table}

\begin{table}[H]
\caption{Yelp Polarized Dataset Results}
\begin{center}
\begin{tabular}{|c|c|c|}
\hline

\textbf{Approach} & \textbf{Number of Bits} & \textbf{Accuracy \%} \\
\hline
Conventional & 6328062  &  89.45\\
\hline
Conventional - Size 2 & 5646338  &  89.45\\
\hline
Conventional - Size 3 & 5077336  & 89.45\\
\hline
Semantic Quantization & 28727  &  80.95\\
\hline
Semantic Compression - 1082 & 19276  &  78.95\\
\hline
\end{tabular}
\label{yelpresults}
\end{center}
\end{table}

\begin{figure}[htbp]
\centering
\includegraphics[width=3.4in]{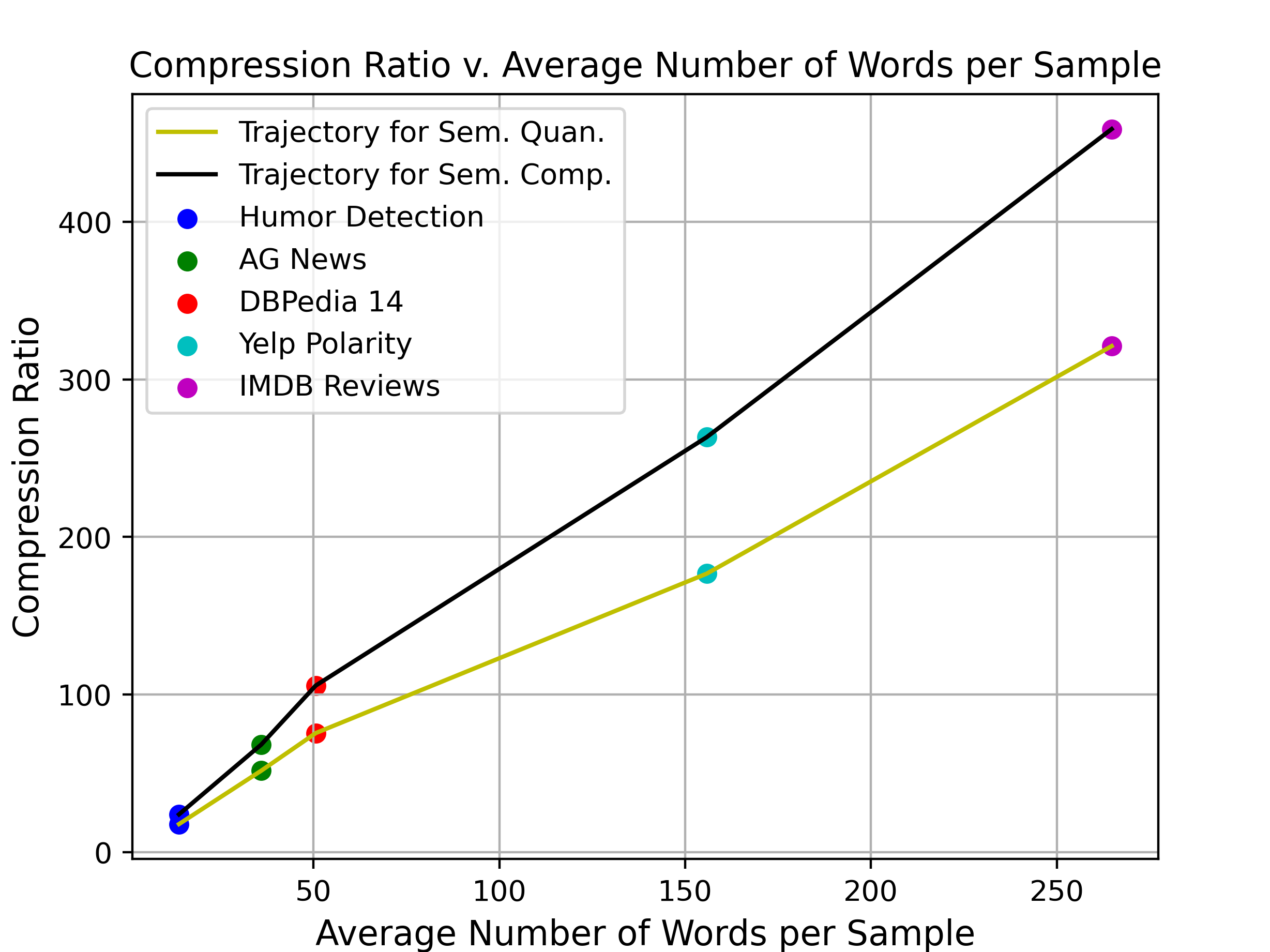}
\caption{Compression ratio versus average number of words per sample.}
\label{fig:comprat_v_avgnumwords} 
\end{figure}

Conducting tests with the same number of samples enables the investigation of average sample length versus compression ratio as given in fig.\ref{fig:comprat_v_avgnumwords} along with compression trajectories. We observe that the efficiency introduced by proposed approaches increases for longer samples. Comparing the trajectories, higher compression ratios are achieved with semantic compression where the gap between approaches increases for larger average lengths per sample.

\subsection{Model Parameters}
\begin{figure}[H]
\centering
\includegraphics[width=2.85in]{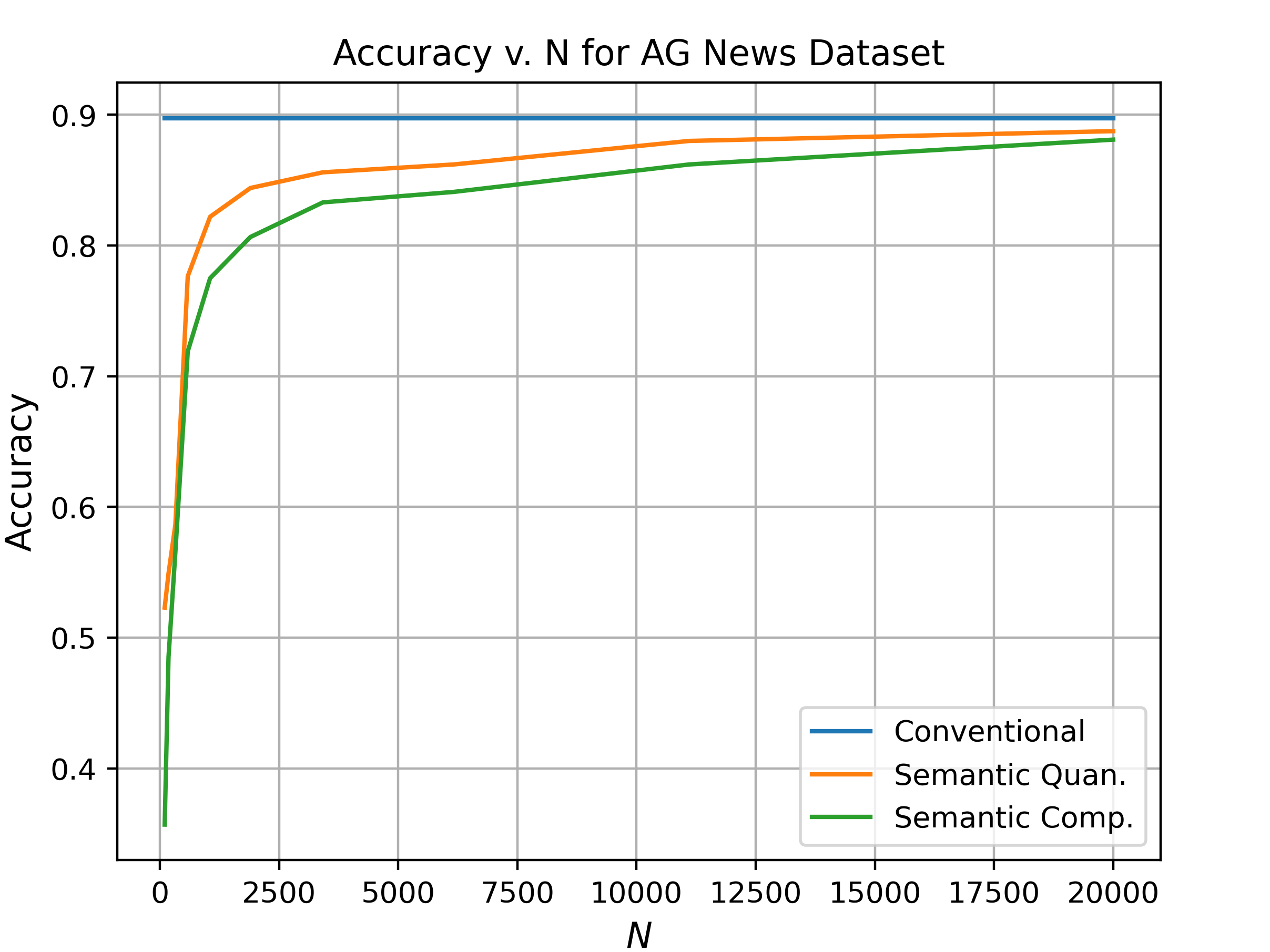}
\caption{Classification accuracy versus \textit{N} value for AG News dataset.}
\label{fig:acc_v_n_agnews} 
\end{figure}

\begin{figure}[H]
\centering
\includegraphics[width=2.85in]{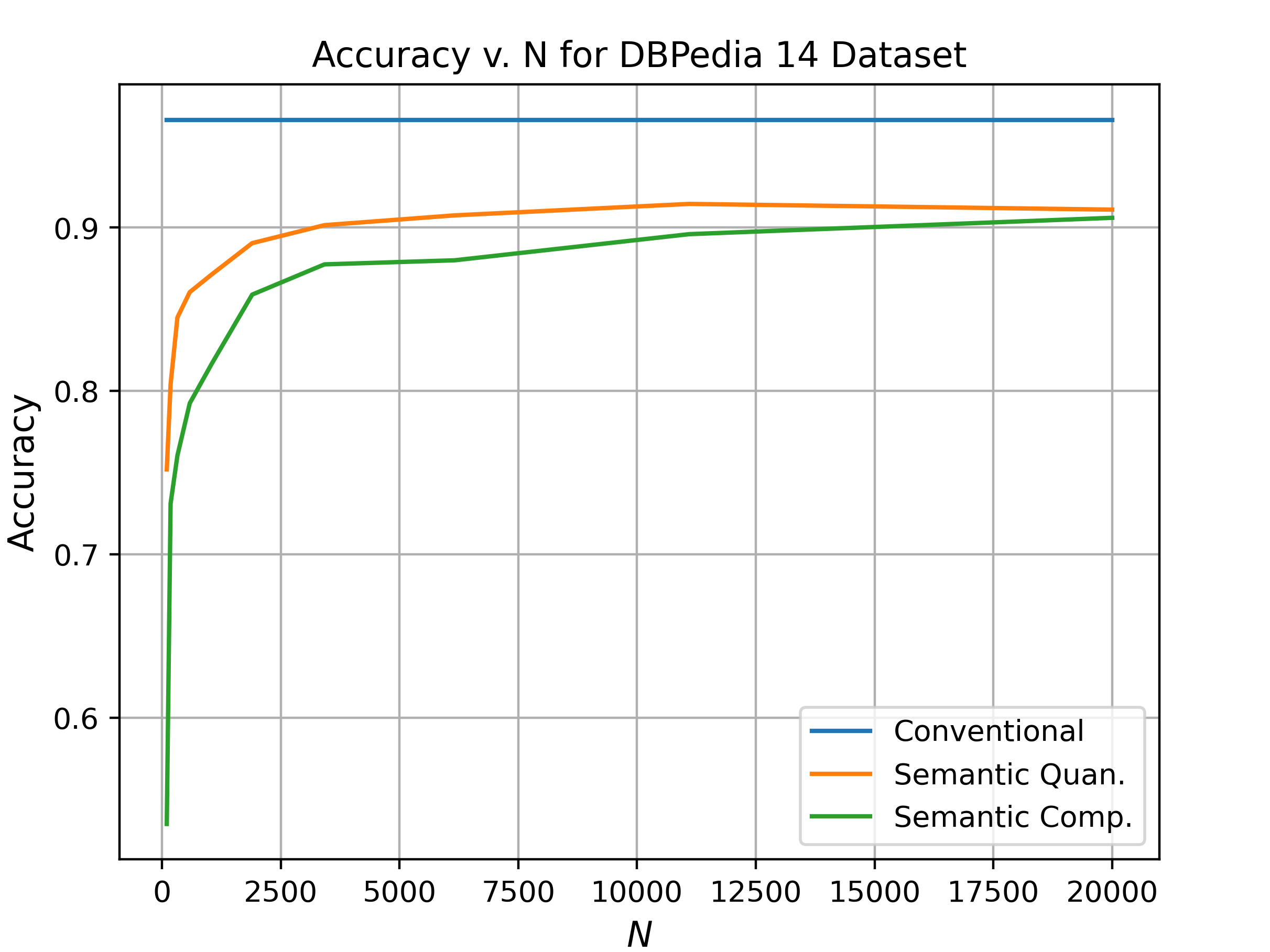}
\caption{Classification accuracy versus \textit{N} value for DBPedia 14 dataset.}
\label{fig:acc_v_n_dbpedia} 
\end{figure}

\begin{math} Q_{pre}^N \end{math} is an \begin{math} \textit{N}  \times  \textit{p} \end{math} dimensional matrix, where \textit{p} is the embedding dimension determined by SBERT neural network. Although large \textit{N} results in a larger matrix requiring storage capacity and computational ability, it should be sufficiently large to provide sufficient samples in each class for semantic index assignment and semantic clustering. We obtain accuracy versus \textit{N} graphs for AG's News and DBPedia 14 datasets in figures \ref{fig:acc_v_n_agnews} and \ref{fig:acc_v_n_dbpedia}. A quick convergence is observed w.r.t \textit{N} in both results, which is beneficial for design complexity-accuracy trade-off by allowing small \textit{N} values. Comparing figures \ref{fig:acc_v_n_agnews} and \ref{fig:acc_v_n_dbpedia}, a faster convergence is achieved in \ref{fig:acc_v_n_dbpedia} although the final error rate is higher w.r.t to the conventional approach. 

We investigate the effect of the pre-trained neural network in SBERT through tables \ref{AGNews_pretrained} and \ref{DBPedia_pretrained}, where accuracies for AG News and DBPedia 14 datasets are given along with dimensions of embeddings obtained by the following pre-trained all-purpose neural networks: i) \textit{all-mpnet-base-v2}; ii) \textit{all-distilroberta-v1}; iii) \textit{all-MiniLM-L12-v2} \cite{sentence-bert}. 

In tables \ref{AGNews_pretrained} and \ref{DBPedia_pretrained}, neural network \textit{mpnet} in SBERT provided superior performance in the majority of obtained results at the cost of higher embedding dimension \textit{p}. Larger \textit{p} value requires a more complex design by resulting in a larger \begin{math} Q_{pre}^N\end{math} matrix. Hence, we state that the design complexity-accuracy trade-off is also valid for pre-trained neural network selection for SBERT.

\begin{table}[H]
\caption{AG News Accuracies - Pre-trained Neural Networks}
\begin{center}
\begin{tabular}{|c|c|c|}
\hline
\textbf{Approach - Pre-trained Neural Network} & \textbf{Acc. \%} & \textbf{Dim. (\textit{p})} \\
\hline
Conventional - \textit{mpnet} & 89.75   & \\
\cline{1-2} 
Semantic Quantization - \textit{mpnet} & 88.75 &   768 \\
\cline{1-2} 
Semantic Compression - \textit{mpnet}  & 88.15  &\\
\hline
Conventional - \textit{distilroberta} & 90.15    & \\
\cline{1-2} 
Semantic Quantization - \textit{distilroberta} & 89.05 &   768 \\
\cline{1-2} 
Semantic Compression - \textit{distilroberta}  & 87.20  &\\

\hline
Conventional - \textit{miniLM-L12} & 89.55  &   \\
\cline{1-2} 
Semantic Quantization - \textit{miniLM-L12} & 87.40 &   384 \\
\cline{1-2} 
Semantic Compression - \textit{miniLM-L12}  & 86.55  &\\
\hline
\end{tabular}
\label{AGNews_pretrained}
\end{center}
\end{table}

\begin{table}[H]
\caption{DBPedia 14 Accuracies - Pre-trained Neural Networks}
\begin{center}
\begin{tabular}{|c|c|c|}
\hline
\textbf{Approach - Pre-trained Neural Network} & \textbf{Acc. \%} & \textbf{Dim. (\textit{p})} \\
\hline
Conventional - \textit{mpnet} & 96.60  &   \\
\cline{1-2} 
Semantic Quantization - \textit{mpnet} & 91.10 & 768 \\
\cline{1-2} 
Semantic Compression - \textit{mpnet}  & 90.60  &\\
\hline
Conventional - \textit{distilroberta} & 96.45    & \\
\cline{1-2} 
Semantic Quantization - \textit{distilroberta} & 90.07 &  768 \\
\cline{1-2} 
Semantic Compression - \textit{distilroberta}  & 88.45 &\\
\hline
Conventional - \textit{miniLM-L12} & 96.30    & \\
\cline{1-2} 
Semantic Quantization - \textit{miniLM-L12} & 88.30 &   384 \\
\cline{1-2} 
Semantic Compression - \textit{miniLM-L12}  & 87.05  &\\
\hline
\end{tabular}
\label{DBPedia_pretrained}
\end{center}
\end{table}

\section{Conclusion}
\label{sec:conclusion}
We have investigated the utilization of semantics for resource savings while preserving the generalizability of conventional models. We have introduced a semantic distortion metric and used it as the objective to propose semantic quantization and compression approaches where semantics are extracted by sentence embeddings. Our results indicate that proposed approaches provide significant efficiency improvement at the expense of a modest decrease in task accuracy. Further directions include the application to different tasks like document retrieval and question-answering.

\bibliographystyle{IEEEtran}{}
\bibliography{semantic-comm-refs}

\begin{thebibliography}{10}
\providecommand{\url}[1]{#1}
\csname url@samestyle\endcsname
\providecommand{\newblock}{\relax}
\providecommand{\bibinfo}[2]{#2}
\providecommand{\BIBentrySTDinterwordspacing}{\spaceskip=0pt\relax}
\providecommand{\BIBentryALTinterwordstretchfactor}{4}
\providecommand{\BIBentryALTinterwordspacing}{\spaceskip=\fontdimen2\font plus
\BIBentryALTinterwordstretchfactor\fontdimen3\font minus \fontdimen4\font\relax}
\providecommand{\BIBforeignlanguage}[2]{{%
\expandafter\ifx\csname l@#1\endcsname\relax
\typeout{** WARNING: IEEEtran.bst: No hyphenation pattern has been}%
\typeout{** loaded for the language `#1'. Using the pattern for}%
\typeout{** the default language instead.}%
\else
\language=\csname l@#1\endcsname
\fi
#2}}
\providecommand{\BIBdecl}{\relax}
\BIBdecl

\bibitem{6gsurvey}
W.~Jiang, B.~Han, M.~A. Habibi, and H.~D. Schotten, ``The road towards 6g: A comprehensive survey,'' \emph{IEEE Open Journal of the Communications Society}, vol.~2, pp. 334--366, 2021.

\bibitem{6gandbeyond}
\BIBentryALTinterwordspacing
``6g and beyond.'' [Online]. Available: \url{https://6gandbeyond.osu.edu/}
\BIBentrySTDinterwordspacing

\bibitem{Shannon}
C.~E. Shannon, ``A mathematical theory of communication,'' \emph{Bell System Technical Journal}, vol.~27, pp. 379--423, 625--656, July, Oct. 1948.

\bibitem{beyondtransmittingbits}
D.~Gündüz, Z.~Qin, I.~E. Aguerri, H.~S. Dhillon, Z.~Yang, A.~Yener, K.~K. Wong, and C.-B. Chae, ``Beyond transmitting bits: Context, semantics, and task-oriented communications,'' \emph{IEEE Journal on Selected Areas in Communications}, vol.~41, no.~1, pp. 5--41, 2023.

\bibitem{guler2014semantic}
B.~Guler and A.~Yener, ``Semantic index assignment,'' in \emph{IEEE Int. Conf. on Pervasive Comp. and Comm. Workshops (PerCom)}, 2014, pp. 431--436.

\bibitem{semcom_game}
B.~Güler, A.~Yener, and A.~Swami, ``The semantic communication game,'' \emph{IEEE Transactions on Cognitive Communications and Networking}, vol.~4, no.~4, pp. 787--802, 2018.

\bibitem{guler2013_semanticcompression}
B.~Guler, A.~Yener, and P.~Basu, ``A study of semantic data compression,'' in \emph{2013 IEEE Global Conference on Signal and Information Processing}, 2013, pp. 887--890.

\bibitem{guler2017_entropy}
B.~Guler, A.~Yener, P.~Basu, and A.~Swami, ``Two-party zero-error function computation with asymmetric priors,'' \emph{Entropy}, vol.~19, no.~12, 2017.

\bibitem{toshea_dl}
T.~O’Shea and J.~Hoydis, ``An introduction to deep learning for the physical layer,'' \emph{IEEE Transactions on Cognitive Communications and Networking}, vol.~3, no.~4, pp. 563--575, 2017.

\bibitem{zhijin_deepsc}
H.~Xie, Z.~Qin, G.~Y. Li, and B.-H. Juang, ``Deep learning enabled semantic communication systems,'' \emph{IEEE Transactions on Signal Processing}, vol.~69, pp. 2663--2675, 2021.

\bibitem{sentence-bert}
N.~Reimers and I.~Gurevych, ``Sentence-bert: Sentence embeddings using siamese bert-networks,'' in \emph{Proceedings of the 2019 Conference on Empirical Methods in NLP}.\hskip 1em plus 0.5em minus 0.4em\relax Association for Computational Linguistics, 11 2019.

\bibitem{affinity_prop}
B.~J. Frey and D.~Dueck, ``Clustering by passing messages between data points,'' \emph{Science}, vol. 315, no. 5814, pp. 972--976, 2007.

\bibitem{bert}
J.~Devlin, M.-W. Chang, K.~Lee, and K.~Toutanova, ``{BERT}: Pre-training of deep bidirectional transformers for language understanding,'' in \emph{Proceedings of the 2019 Conference of the North {A}merican Chapter of the Association for Computational Linguistics}.\hskip 1em plus 0.5em minus 0.4em\relax Minneapolis, Minnesota: Association for Computational Linguistics, Jun. 2019, pp. 4171--4186.

\bibitem{facenet_triplets}
F.~Schroff, D.~Kalenichenko, and J.~Philbin, ``Facenet: A unified embedding for face recognition and clustering,'' in \emph{2015 IEEE Conference on Computer Vision and Pattern Recognition (CVPR)}, 2015, pp. 815--823.

\bibitem{text_classification_dataset_lecun}
X.~Zhang, J.~Zhao, and Y.~LeCun, ``Character-level convolutional networks for text classification,'' in \emph{Advances in Neural Information Processing Systems}, C.~Cortes, N.~Lawrence, D.~Lee, M.~Sugiyama, and R.~Garnett, Eds., vol.~28.\hskip 1em plus 0.5em minus 0.4em\relax Curran Associates, Inc., 2015.

\bibitem{dataset_humor}
I.~Annamoradnejad and G.~Zoghi, ``Colbert: Using bert sentence embedding for humor detection,'' \emph{arXiv preprint arXiv:2004.12765}, 2020.

\bibitem{imdb_dataset}
A.~L. Maas, R.~E. Daly, P.~T. Pham, D.~Huang, A.~Y. Ng, and C.~Potts, ``Learning word vectors for sentiment analysis,'' in \emph{Proceedings of the 49th Annual Meeting of the Association for Computational Linguistics: Human Language Technologies}.\hskip 1em plus 0.5em minus 0.4em\relax Portland, Oregon, USA: Association for Computational Linguistics, June 2011, pp. 142--150.

\bibitem{UMAP}
L.~{McInnes}, J.~{Healy}, and J.~{Melville}, ``{UMAP: Uniform Manifold Approximation and Projection for Dimension Reduction},'' \emph{ArXiv e-prints}, Feb. 2018.

\end{thebibliography}
\end{document}